# Petabyte Scale Data Mining: Dream or Reality?

Alexander S. Szalay, Johns Hopkins University
Jim Gray, Microsoft Research
Jan vandenBerg, Johns Hopkins University

# Petabyte Scale Data Mining: Dream or Reality?

Alexander S. Szalay[a], Jim Gray[b] and Jan Vandenberg[a]

[a]Department of Physics and Astronomy, The Johns Hopkins University, Baltimore, MD 21218
[b]Microsoft Research, San Francisco, CA 94105

## ABSTRACT

Science is becoming very data intensive[1]. Today's astronomy datasets with tens of millions of galaxies already present substantial challenges for data mining. In less than 10 years the catalogs are expected to grow to billions of objects, and image archives will reach Petabytes. Imagine having a 100GB database in 1996, when disk scanning speeds were 30MB/s, and database tools were immature. Such a task today is trivial, almost manageable with a laptop. We think that the issue of a PB database will be very similar in six years. In this paper we scale our current experiments in data archiving and analysis on the Sloan Digital Sky Survey[2,3] data six years into the future. We analyze these projections and look at the requirements of performing data mining on such data sets. We conclude that the task scales rather well: we could do the job today, although it would be expensive. There do not seem to be any show-stoppers that would prevent us from storing and using a Petabyte dataset six years from today.

## 1. COLLECTING PETABYTE DATASETS

### 1.1 Generating a Petabyte

There are several concrete experiments (LSST[4], PAN-STARRS[5]), which are aiming at such data sets. Instead of taking the precise parameters of either, we will try do define an idealized experiment that is in the same ballpark, with the approximate capability of generating a Petabyte of imaging data per year. We are working backwards from a 1 Petabyte/year data size. Assuming pixel sizes of about 0.25", the whole sky is about 10 Terapixels. We take this our canonical observation. This can be split as half the sky in two bands, or quarter of the sky in four bands, etc, all yielding about 20TB of raw imaging data, considering 2 bytes per pixel. It is unlikely that we will attempt to image the sky in 50 bands, thus we assume that the primary aim of the experiment is to get time-domain data, detecting transients and moving objects, in a small number of filters.

An image from a 5 Gigapixel camera is 10 Gbytes. Taking a series of 1 minute exposures translate to about 5 TB/night/Gpixels. The proposed experiments plan on covering the sky in about 4 nights, this translates to about 5TB/night when the weather is good enough for imaging, a consistent number. We also assume an excellent site, with 200 nights of observations per year, yielding 50 passes over the appropriate area of the sky, generating a Petabyte of imaging data. We went through this exercise to ensure that the basic assumptions about the volume and the rate of the data are correct, and are in line with the tentative plans of the relevant projects.

### 1.2 Processing a Petabyte

The incoming data stream is 5TB/8 hrs = 170MB/s. It would be a challenge to write such a data stream to tapes. Writing such a data stream to disks is easy, even today. Given the trend in disk prices we assume that tape storage will be entirely obsolete in six years. In the Sloan Digital Sky Survey running the photometric pipelines on a 6MB image takes about 10 seconds on a typical single processor 2.2GHz Linux workstation. Taking this 0.6MB/s processing rate, we would need about 300 processors to perform the analysis in essentially real time today. Moore's Law predicts a 16-fold speedup in processing power over the next six years. We assume that the processing tasks may become somewhat more complex, but this will be compensated by incrementally better image processing algorithms. It appears that of the order of 20 processors should be able to keep up with the tasks of processing the incoming data stream.

### 1.3 Storing a Petabyte

We assume that we will want to store the Petabyte, at least over a year then possibly discard. Today's prices for a TB of commodity storage are around $2K/TB. The trend has been a factor of 2 decrease in price per year. A PB storage would cost today about $2M, but if the trend holds, this would fall to about $32K in six years. Even if this is a bit extreme, disk costs will not be a major issue. Disk sizes are approximately doubling per year; in about 2 years we should see 0.5TB disks, growing to a size between 1 to 4TB in six years. We can deal with farms consisting of 500-1000 disks today without much difficulty. Assuming that the disk sequential IO speeds will only grow very mildly, roughly proportional to the square root of the size, to about 150MB/s/disk, we should still be able to read the data at an aggregate rate of 75GB/s: we can scan through all the images and read them off the disks in less than 4 hours.

### 1.4 Economics

Our idealized survey that gathers a Petabyte each year involves hundreds of people, a multi-million dollar telescope budget, and huge intellectual investment. In round numbers it is a $100M project (for the data gathering) with a $25M for software. The yearly burn rate is $5M for the observatory and the pipeline.

Now consider the hardware economics. Suppose that the data arrives at a Petabyte a year. In year 1 we need the 20 servers for the pipeline, we need the 500 disks for the bandwidth and we need some processing power to analyze the data. In today's dollars this would be a few hundred thousand dollars of equipment.

In year N, we need N times more disk bandwidth and N times more storage, but the data rate is the same. So, if Moore's law holds true after 3 years we need 1/4 the processors for the pipeline (because processors are 3x faster), we need about the same number of processors for the analysis (they are 4x faster but there is 3x more data,) and we need 2x more disks (disks have gotten a little faster, but we need to go through 3x more data). The net of this is that the hardware budget probably stays constant or even shrinks during the lifetime of the project, even though the data volumes grow enormously. And the total hardware budget is a tiny fraction of the project budget or even of the software budget.

## 2. DATA ORGANIZATION

### 2.1 Data Layout

For scientific analysis we will use the output of the processing pipelines, stored in a database. The stored data products will consist of object catalogs, various calibration parameters, observing logs, email archives, source code of the software, and any and all information related to how the data was derived. The dominant part will be the object catalog. Projecting from the SDSS[2], we expect deeper exposures, and a higher surface density of objects, probably by a factor of 10, yielding about 2 billion objects / pass. Taking 50 passes per year the expected catalog size is 100 billion objects. Considering 1KB/object, the resulting database will be about 100TB, 10% of the size of the raw pixel data, a number similar to the SDSS.

We will need to create several indices, in particular to perform spatial queries. In order to build chains of objects at the same location we will need to find objects near one another. In order to avoid coordinate singularities, we have built a class library with an SQL API for the SDSS archive that can perform efficient spatial searches within a spherical polygon. In the SDSS database we have created a precomputed Neighbors table, which contains pairs of objects within a predefined angular distance. We need also indices built on the time-domain behavior of objects. One should consider a 20% overhead for indices.

Any form of scientific analysis will need to access the object catalog. Proper data organization can be much help. Objects can be classified as static (both space and time), variable (same location, varying flux), transient, moving, defects, etc. We will clearly need to store each observation/detection of an object. But, it will also be useful to create a summary, or 'master' copy of each unique object location, than can then be linked to the chain of detections. Many of the analyses and queries will first look at the summary, in particular if that already contains a precomputed time-domain classification and possibly a light-curve. Since the catalog will be the aggregation of 50 scans over the data, the master catalog will be 50 times smaller, thus any query that does not have to look at all the detections can run 50 times faster. After accounting for some overhead for the summary, the size of the master catalog can be reduced by a factor of 30, to about 4TB.

In a similar fashion, we will also want to store a 'best' sky, which is coadded from all the images taken at on point in the sky. Taking our canonical 10 Terapixels, and assuming 2 bytes for the image and 1 byte for the noise, this becomes 30 TB. Of course, this only captures the static, 'mean' sky. If we consider that about 1% of the sky contains variable pixels, and we would like to save those for each pass, the total size of the extracted images will be 30x(1+0.5) = 45 TB.

**2.2 Database loading**

Database loading should be able to keep up with the processing rate. Since the database size is about 12% of the imaging data, the peak data rate for loading is about 20MB/s, with an average load speed less than 5MB/s. This is easily achievable today. The most CPU intensive part would be the index calculations. The expected speedup of CPUs will turn this into a trivial task. If we need to reload the database from scratch, the task is a little more challenging. What would it take to load the database in two weeks? 120TB in two weeks is about 125MB/s, mostly limited by CPU. We could still do it today, but we would need to load data in parallel. The access requirements for the hardware architecture already dictate an 8-way parallel database architecture, and such a system would make the loading task easy: 15MB/s/brick.

**2.3 Backup, Recovery and Replication**

The backup and replication of such a data set is a non-trivial question. In order to avoid disaster from simple disk failures is to configure the database over a set of mirrored disks, in a RAID10 configuration, currently much preferred over a RAID5 due to its better sequential performance. Probably the best backup strategy is to keep a copy at a separate geographic location. Replicating a total of 165TB of data in one go over the network can be rather time consuming. A typical academic site, like an observatory can at most expect an OC-3 link today, with a bandwidth of 155Mbits/s. With a 65% utilization, this translates to about 10MB/s, our database can be transferred in 200 days. The current costs of a dedicated OC-3 line are rather high, turning such a solution unaffordable.

A modern version of the 'sneakernet'[6] may be more feasible: with 4TB disks, 8 disks per box, one can build compact 'transfer-bricks' that can keep 32TB, albeit at a relatively slow scanning speed. The cost of such a box should be the same as our canonical $2K 'brick' today. Shipping 5 of these via overnight mail, then point-to-point copying the contents over using dedicated 10 Gbit/sec links between the transfer bricks and the database bricks, running in parallel will enable us to move data within two days from one site to another. The transfer bricks can be reused many times, like shipped on to the next mirror site.

**2.4 Sequential Scans and Hardware Architecture**

The worst-case scenario in any simple analysis is sequential scan: we search and filter on an attribute that is not contained by any of the indices. The speed of sequential IO of servers has been improving dramatically over the last few years. Our own measurements with inexpensive Intel 'bricks' have given us 10MB/s in 1995, 30MB/s in 1997, 200MB/s in 2000, and 400MB/s this year. It is fairly reasonable to assume that a single database 'brick' will be able to scan at around 4GB/s in six years. Assuming again 150MB/s per disk, each server needs to have about 30 disks.

The limiting factor in the sequential IO will be the total number of disks. If we store our 120TB database on large, 4TB disks, we only need 30 of them, thus a single database server can run the database, but a single scan will be 7.4 hours. Scanning the master will be 20 times faster, about 20 minutes. The only way to increase the speed of the sequential IO is to increase the number of disks, by using smaller disks. If we need to bring the scan time down to 1 hour, we need to use no larger than 500GB disks. This translates to 240 disks, running in parallel. As we have seen, 30 disks will saturate a server, thus we need at least 8 servers running in parallel. With such a configuration we can scan the master database in less than 3 minutes. But, separating the summary database to a single disk is not a good idea: it is the massive parallelism of the disk drives that gives us the fast performance.

One might consider of sharing the disks with the raw image archive. Assigning a small area (%12) of each disk to the catalog database, one would gain in parallelism. In particular, the outer edge of the disks provides about twice as large sequential data rate than the inner parts (disks rotate at a fixed rpm), thus we may get close to the transfer rate of the interface.

## 3. MINING THE DATA

### 3.1 Simple queries, filtering

The basic task the system has to deliver is the ability to provide results from simple queries quickly. For the SDSS SkyServer[7,8] we found that writing down 20 typical scenarios/queries has been very helpful in designing the database layout, and index structure. Having 20 different astronomy 'patterns' was large enough that no single index was able to cover, but not large enough that one could not run various tests easily during the design and test phase. We will also need to give access to image cutouts, and meaningful time-domain information, like light-curves, or estimated trajectories of moving objects, like asteroids, etc.

### 3.2 Real-Time Triggers

In time-domain astronomy one of the most interesting things to do is to try to detect transient objects, like supernovae very quickly, so that they can be followed through their history, possibly with other telescopes. In order to perform the detections as fast as possible, we will assume, that at least a first-pass object detection will be performed at the telescope. For comparison, we will need to stream the predicted object catalog through, synchronized with the observations, so that we can do on-line detection of a transient. A finer, proper photometric reduction can be done later. The data stream in this case is expected to be quite trivial, a few MB/sec.

An alternative is to stream the `summary' version of the sky against the catalog, plus a designation of known sources, and do the transient detection by subtracting the images and only process the residuals. This requires a considerably faster archival data stream, essentially equivalent to the incoming data stream itself. Unless the summary images are stored at the telescope site, this may not be practical, due to network throughput requirements.

### 3.3 Statistics

Focusing on time-domain data, one of the most important thing is to build first timelines of variable and transient objects and identify moving objects. For the variable objects we need to model the variability, and classify them to periodic variables and transients, and try to identify their types. We will need to fit model lightcurves to irregularly sampled data. This process is reasonably well understood.

For moving objects, we need to map their detections on possible models for orbits, separating asteroids from space debris, and then search in this parameter space for multiple detections of the same object, identified by their proximity in the orbit parameters. The models can be refined as new detections are added.

Spatial clustering measures, in particular correlation functions have been one of the most frequently used statistical tool in astrophysics. The naïve algorithms have rather poor scaling properties: $O(N^2)$ for two point, $O(N^M)$ for an M-point correlation function. Recent techniques of tree-based codes give an $N^{3/2}$ scaling[9], irrespectively of the rank, and grid based techniques[10] scale between NlogN to $N^{3/2}$.

### 3.4 Discovering Patterns

There are several aspects of data mining, which will all be important for such data sets. The main problem is that most data mining algorithms, which involve cluster finding, are based on some distance metric. Thus, they are inherently scaling faster than linear in the number of objects. Our premise is that no algorithm worse than NlogN will be feasible in the world of Petabytes. The number of objects we can detect will roughly scale with detector sizes, which in turn will scale with Moore's law, the same rate as our processing power grows. For a billion objects logN is 30, we can compensate for such factors with an increased level of parallelism, but we could not deal with an $N^2$ algorithm.

In particular, outlier detection has been rather computer intensive in the past. EM clustering is one of the better algorithms used in this context. As shown by A. Moore and collaborators, by preorganizing data into a multidimensional kd-tree, and only computing distances for objects near in the tree sense, rejecting distant pairs, one can achieve a close to NlogN behavior, the dominant cost being the presorting[11].

We will need a variety of supervised and unsupervised classification techniques, matching objects detected in multiple bands to spectral templates, and derive photometric redshifts, and rest-frame parameters, if at all possible. Gravitational lensing manifests itself in distortions of images of distant galaxies, sometimes so weak that it can only be detected in the statistical sense, through correlation functions. We will need better parameterizations of shapes, possibly in terms of rotationally invariant quantities. Using Minkowski functionals and valuations[12] looks like a promising direction.

## 4. SUMMARY

In this paper we have tried to scale our current approach on building a general-purpose archive for the Sloan Digital Sky Survey six years into the future, when Petabyte datasets are expected to be the norm. Our technique involves relational databases, where we are trying to scale-out, by using multiple inexpensive 'bricks', rather than scale-up to large, multiprocessor 'mainframes' and storage network arrays. The only bottleneck we found is achieving the required sequential IO performance – it requires a large number of disks. By interleaving the image storage with the catalog storage one can reach stunning a stunning performance over 500+ disks: a scanning speed of over 75GB/sec. Scanning the 4TB summary catalog at this rate would only take 53 seconds!

Our experience has shown that both from the perspective of raw performance and total price of ownership, such a scale-out is a much better strategy than buying expensive storage arrays. The raw sequential IO of direct attached disk drives over multiple controllers is considerably faster than expensive FiberChannel arrays. Given the trends in processor power, disk capacity and prices, the total price of the hardware necessary is expected to be rather affordable by the time it is needed.

Based upon the SDSS experience, if one has aa careful processing plan and a good database design, one can create a small 'summary' catalog, which will cover most of the scientific uses, the rest of the data needs to be looked at much less frequently. The small size of the summary catalog enables it to be offered even through a website (the Jan 2003 data release of SDSS will already contain over 1TB catalog data), it can be replicated at many sites, it can even be updated regularly at those mirror sites, and we can also create many custom views, for particular science uses much easier. At the same time all the data necessary to reprocess either the summary or the base catalog data can also be saved. It is important that the provenance of the data be kept.

Beyond the first year the size of the archive of such a project will be growing at a linear rate, while the technology keeps evolving at an exponential rate: the first year of the project will be the hardest!

## ACKNOWLEDGEMENTS

AS acknowledges support from grants NSF AST-9802 980, NSF KDI/PHY-9980044, NASA LTSA NAG-53503 and NASA NAG-58590. He is also supported by a grant from Microsoft Research. Our effort has also been generously supported with hardware by Intel, and Compaq/HP.